\documentclass[pra,twocolumn,showpacs,superscriptaddress,byrevtex]{revtex4}

\usepackage{graphicx}
\usepackage{dcolumn}
\usepackage{amsmath}
\usepackage{epsfig}
\usepackage{bm}
\usepackage{amssymb}

\usepackage[dvips]{hyperref}
\hypersetup{colorlinks=true,linkcolor=blue,citecolor=blue,urlcolor=black}

\newcommand{\ket}[1]{\ensuremath{|\,{#1}\,\rangle}}
\newcommand{\bra}[1]{\ensuremath{\langle\,{#1}\,|}}

\begin{document}

\title{Optimal measurement bases for Bell-tests based on the CH-inequality}

\date{\today}

\author{G. Lima}
\email{glima@udec.cl}
\affiliation{Departamento de F\'{i}sica, Universidad de Concepci\'{o}n, 160-C Concepci\'{o}n, Chile}
\affiliation{Center for Optics and Photonics, Universidad de Concepci\'{o}n, Chile}
\affiliation{MSI-Nucleus on Advanced Optics, Universidad de Concepci\'{o}n, Chile}
\author{E. B. Inostroza}
\affiliation{Departamento de F\'{i}sica, Universidad de Concepci\'{o}n, 160-C Concepci\'{o}n, Chile}
\affiliation{Center for Optics and Photonics, Universidad de Concepci\'{o}n, Chile}
\author{R. O. Vianna}
\affiliation{Departamento de F\'{i}sica, ICEx, Universidade Federal de Minas Gerais, Belo Horizonte, MG, Brazil}
\author{J.-{\AA}. Larsson}
\affiliation{Institutionen f\"or Systemteknik, Link\"opings Universitet, SE-58183 Link\"oping, Sweden}
\author{C. Saavedra}
\affiliation{Departamento de F\'{i}sica, Universidad de Concepci\'{o}n, 160-C Concepci\'{o}n, Chile}
\affiliation{Center for Optics and Photonics, Universidad de Concepci\'{o}n, Chile}



\begin{abstract}
The Hardy test of nonlocality can be seen as a particular case of the Bell tests based on the Clauser-Horne (CH) inequality. Here we stress this connection when we analyze the relation between the CH-inequality violation, its threshold detection efficiency, and the measurement settings adopted in the test. It is well known that the threshold efficiencies decrease when one considers partially entangled states and that the use of these states, unfortunately, generates a reduction in the CH-violation. Nevertheless, these quantities are both dependent on the measurement settings considered, and in this paper we show that there are measurement bases which allow for an optimal situation in this trade-off relation. These bases are given as a generalization of the Hardy measurement bases, and they will be relevant for future Bell tests relying on pairs of entangled qubits.
\end{abstract}

\maketitle


\section{Introduction}

In 1992, Hardy \cite{Hardy1} showed that an experiment with electrons
and positrons could be used to test nonlocality when
certain types of joint measurements are considered. Later his
observations were presented in a theorem-like form, that holds
for pure systems of two non-maximally entangled qubits (``quantum bits''), regardless
of the degree of freedom used to encode the qubits
\cite{Hardy2,Goldstein}. The logic of Hardy's argument, which is discussed in more detail in \cite{Mermin}, can be described as follows: suppose that one can perform joint measurements in this composite system, and that the measuring
settings of the two apparatuses are denoted by the
parameters $\theta_1$, $\theta_2$, $\theta_3$ and $\theta_4$.
Consider also the following probabilities of joint detection: i)
$P(\tilde{\theta_1},\theta_3)$, ii) $P(\theta_1,\theta_4)$ and iii)
$P(\theta_2,\tilde{\theta_4})$, where $\tilde{\theta_1}$ denotes the
orthogonal configuration to the setting defined by $\theta_1$.
For any non-symmetric entangled state, it is always possible to find the parameters $\theta$'s such that (i),
(ii) and (iii) are null and $P(\theta_2,\theta_3)\neq0$. This
creates a contradiction between quantum mechanics and local
theories, for which the fact of having (i), (ii) and (iii) null
implies that $P(\theta_2,\theta_3)=0$. The difference between the
value predicted by quantum mechanics for $P(\theta_2,\theta_3)$ and
the value predicted by local theories is known as ``Hardy
fraction'' \cite{Torgerson,Demartini1,Mataloni,Pino}.

As it was discussed by Mermin and Garuccio in \cite{Mermin,Garuccio}, Hardy's test can be generalized when it is written in terms of the following
inequality
\begin{equation} \label{Garu}
P(\phi_2,\phi_3) \leq P(\tilde{\phi_1},\phi_3) + P(\phi_1,\phi_4) + P(\phi_2,\tilde{\phi_4}),
\end{equation} which holds for any choice of $\phi_1$, $\phi_2$, $\phi_3$ and $\phi_4$, while the
Hardy's test is valid only for special values of $\phi$'s. This inequality is equivalent to the Clauser-Horne (CH) inequality
\cite{Garuccio,ClauserHorne}. The quantum violation of the
CH-inequality is the difference between the value of
$P(\phi_2,\phi_3)$ predicted by quantum mechanics, and the value
given by the sum of the probabilities on the right-hand side of
Eq~(\ref{Garu}), which is the maximum value allowed for
$P(\phi_2,\phi_3)$ by local hidden variable theories \cite{merminLHV}.

In this work we study the dependence of the CH-inequality violation with the measurement bases used in the Bell tests, and we analyze how these measurements affect the required efficiency for closing the detection loophole in these experiments \cite{Santos}. It is well known that the required efficiencies decrease when one works with partially entangled states but, unfortunately, the use of these states generates a reduction in the CH-inequality violation. Nevertheless, the CH-inequality violation and the required efficiency are both dependent on the measurement settings adopted, and here we show that there are measurement bases that allow for an optimal situation in this trade-off relation. These bases are given as a generalization of the Hardy measurement bases.

The relevance of studying such properties of the CH-inequality comes from the fact that it usually outperforms most of the known bipartite Bell inequalities, specially when systems of two entangled qubits are considered \cite{BrunnerPLA}. In fact, as far as we know, for symmetric Bell tests it is only slightly outperformed when high-dimensional entangled systems and multi-setting Bell inequalities are considered \cite{BrunnerPRL10}.

\section{The CH-inequality violation and measurement bases}

To obtain the quantum violation of the CH-inequality for a pure
system of two entangled qubits, we first write the general state of
this system in terms of the Schmidt basis
$\{\ket{\pm}^{(1)},\ket{\pm}^{(2)}\}$ \cite{Schimdt}
\begin{equation} \label{Psi}
\ket{\Psi} = \alpha \ket{+^{(1)},+^{(2)}} + \beta \ket{-^{(1)}, -^{(2)}},
\end{equation} where the coefficients $\alpha$ and $\beta$ are real and positive.

Then, we consider the following general measurement basis
\begin{equation} \label{basis1}
\ket{v^{(k)}}_{\phi} = \sin{\phi} \ket{+^{(k)}} + e^{i \nu_{\phi}} \cos{\phi} \ket{-^{(k)}},
\end{equation} where $k=1,2$ denotes the particles subspaces and the orthogonal vector $\ket{u^{(k)}}_{\phi}$ is given by $\ket{u^{(k)}}_{\phi} = \cos{\phi} \ket{+^{(k)}} - e^{i \nu_{\phi}} \sin{\phi} \ket{-^{(k)}}$.

In terms of this basis, the measurement with the experimental
apparatus in the orientation $\phi$ is represented by the projector
$P^{(k)}_{\phi} = \ket{v^{(k)}}_{\phi} \bra{v^{(k)}}$, and the
measurement $\tilde{\phi}$, which is done with the apparatus in a
configuration orthogonal to $\phi$, is represented by the projector
$P^{(k)}_{\tilde{\phi}} = \ket{u^{(k)}}_{\phi} \bra{u^{(k)}}$. The probabilities for
coincidence detection $P(\phi,\gamma)$, $P(\tilde{\phi},\gamma)$ and
$P(\phi,\tilde{\gamma})$ are given by $|_{\phi} \bra{v^{(1)}}_{\gamma} \bra{v^{(2)}}\ket{\Psi}|^2$, $|_{\phi} \bra{u^{(1)}}_{\gamma} \bra{v^{(2)}}\ket{\Psi}|^2$ and $|_{\phi} \bra{v^{(1)}}_{\gamma} \bra{u^{(2)}}\ket{\Psi}|^2$, respectively. The operator corresponding to the CH-inequality can be written as $\hat{I}_{CH}= (P^{(1)}_{\phi_2} - P^{(1)}_{\tilde{\phi}_1}) \otimes P^{(2)}_{\phi_3} - P^{(1)}_{\phi_1} \otimes P^{(2)}_{\phi_4} - P^{(1)}_{\phi_2} \otimes P^{(2)}_{\tilde{\phi}_4}$.

\begin{figure}[t]
\begin{center}
\rotatebox{270}{\includegraphics[width=0.35\textwidth]{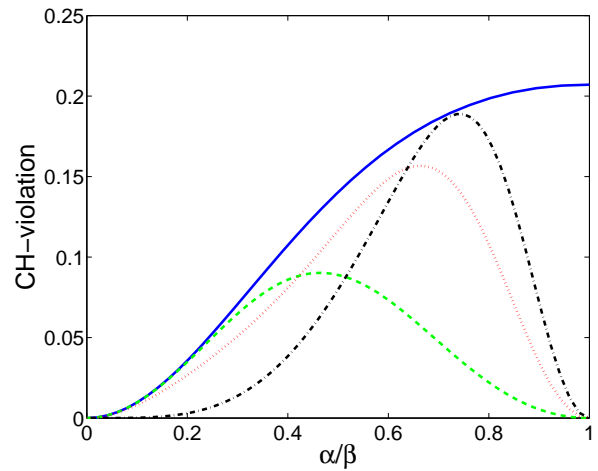}}
\end{center}
\vspace{-0.5cm}
\caption{ (Color online) The CH-inequality violation. The dashed curve (green curve) is obtained when one considers the Hardy measurements bases given in Eq~(\ref{Harbasis}). The dotted curve (red curve) and the dashed-dotted one (black curve) are obtained when the measurements are done with the experimental apparatuses in the orientations defined by the generalized Hardy measurement bases defined by Eq.~(\ref{basis1}) and Eq.~(\ref{newbases}) with $n=1; m=7$ and $n=3; m=10$, respectively. The solid curve (blue curve) is the maximum allowed for the CH-violation for a given value of $\alpha/\beta$.} \label{CHVio}
\end{figure}

When the measurements done in the experiment are characterized by the states $\ket{v^{(1)}}_{\phi_1}$, $\ket{v^{(1)}}_{\phi_2}$, $\ket{v^{(2)}}_{\phi_3}$ and $\ket{v^{(2)}}_{\phi_4}$ given in the particular form of
\begin{eqnarray} \label{Harbasis}
\ket{v^{(1)}}_{\phi_1} &= &\frac{\beta^{\frac{1}{2}} \ket{+^{(1)}} - \alpha^{\frac{1}{2}} \ket{-^{(1)}}}{\sqrt{\alpha + \beta}}, \nonumber\\
\ket{v^{(1)}}_{\phi_2} &= &\frac{\beta^{\frac{3}{2}} \ket{+^{(1)}} + \alpha^{\frac{3}{2}} \ket{-^{(1)}}}{\sqrt{\alpha^3 + \beta^3}},\nonumber\\
\ket{v^{(2)}}_{\phi_3} &= &\frac{\beta^{\frac{3}{2}} \ket{+^{(2)}} - \alpha^{\frac{3}{2}} \ket{-^{(2)}}}{\sqrt{\alpha^3 + \beta^3}},\nonumber\\
\ket{v^{(2)}}_{\phi_4} &= &\frac{\beta^{\frac{1}{2}} \ket{+^{(2)}} + \alpha^{\frac{1}{2}} \ket{-^{(2)}}}{\sqrt{\alpha + \beta}},
\end{eqnarray} we get that $P(\tilde{\phi_1},\phi_3) = P(\phi_1,\phi_4) = P(\phi_2,\tilde{\phi_4}) = 0$ and that $P(\phi_2,\phi_3) = (\frac{\alpha \beta(\alpha-\beta)}{1-\alpha \beta})^2$, which is exactly the fraction deduced by Hardy in Ref \cite{Hardy2}. It has the well known maximum value of approximately $9\%$ when $\alpha / \beta \approx 0.46$. We refer to Eq.~(\ref{Harbasis}) as Hardy measurement bases, and the curve for the Hardy fraction is plotted in Fig.~\ref{CHVio} as a function of $\alpha / \beta$. This parameter $\alpha / \beta$ is directly linked with the concurrence of the state of Eq.~(\ref{Psi}), and it has been widely used for studying Hardy's proof of nonlocality \cite{Hardy2,Demartini1,Mataloni,Pino}. Obviously, a maximally entangled state has $\alpha / \beta =1$, and the degree of entanglement of Eq.~(\ref{Psi}) decreases whenever $\alpha / \beta \rightarrow 0$.

Now, it is interesting to note that by considering rather similar measurements bases one can already see important changes in the CH-inequality violation curve. We can obtain a generalization of the previous case [Eq~(\ref{Harbasis})], by doing the values of the sines and cosines of the measurement bases [Eq~(\ref{basis1})] as
\begin{eqnarray} \label {newbases}
\sin{\phi_1} &= &\sin{\phi_4} = \frac{\beta^{\frac{n}{2}}}{\sqrt{\alpha^n + \beta^n}}, \nonumber \\
\cos{\phi_1} &= &-\cos{\phi_4} = -\frac{\alpha^{\frac{n}{2}}}{\sqrt{\alpha^n + \beta^n}}, \nonumber \\
\sin{\phi_2} &= &\sin{\phi_3} = \frac{\beta^{\frac{m}{2}}}{\sqrt{\alpha^m + \beta^m}}, \nonumber \\
\cos{\phi_2} &= &-\cos{\phi_3} = \frac{\alpha^{\frac{m}{2}}}{\sqrt{\alpha^m + \beta^m}},
\end{eqnarray} and $\nu_{\phi_1} = \nu_{\phi_2} = \nu_{\phi_3} = \nu_{\phi_4} = 0$ with $n \neq m$. In this case, we obtain curves which have distinct values for their maximum and that are maximized for different entangled states. The curves for the new measurement bases defined by $n=1$, $m=7$ and for $n=3$, $m=10$ are also plotted in Fig.~\ref{CHVio}. For this last case, the fraction of the pair of photons that violates the local realism have a maximum of $18.8\%$ when $\alpha / \beta = 0.74$.

The maximum of the CH-inequality violation for each entangled state (for each value of $\alpha / \beta$) can be obtained numerically by means of the well known Conjugate Gradient (CG) method \cite{Press92}. For doing this, we first considered an eight variable function
\begin{equation}
B(\{\phi_i\},\{\nu_i\}) = \langle \Psi \vert \hat{I}_{CH} \vert \Psi \rangle
\label{eq:Bell_function}
\end{equation} with $i=1,2,3,4$ defining the parameters of the operator $\hat{I}_{CH}$. The heuristic of the CG method is to use the local gradient, in a point of the parameter space (the space defined by the variables $\{\phi_i\}$ and $\{\nu_i\}$ ), to reach the closest maximum point (for finding a minimum, the target function is multiplied by -1). The algorithm converges when the gradient is null. To map all the local maxima, and decide which is the global maximum, we run the CG method for a large uniform sample of points in the parameter space. In order to certify that a global maximum has been reached, for each $\alpha/\beta$, we ran the CG method for samples of sizes of $10^3$, $10^4$, and $10^5$. The solid curve (blue curve) shown in Fig.~\ref{CHVio}, is the maximal CH-violation allowed for a given value of $\alpha / \beta$. We can see that it approximates to $20,7\%$ ($\frac{1}{\sqrt{2}} - \frac{1}{2}$) when the degree of entanglement increases \cite{Wootters}. This bound in the quantum violation of the Clauser-Horne inequality, or of the equivalent inequality of Clauser, Horner, Shimony and Holt \cite{CHSH}, is well known and its existence was first deduced in Ref \cite{Cirelson}.

\section{The threshold detection efficiency and measurement bases}

The original Bell inequality \cite{Bell} is a constraint on the correlations of the measurements that can be performed on a composite system. It is obeyed by any local and deterministic description used for the system and the measurement apparatuses, and was deduced by assuming certain types of measurement results. It seems to be unsuitable to account for the inefficiencies of the detectors and the noisy background surrounding the experiment. Nevertheless, it is important to note that there is a generalization of this type of limit which accounts for detection efficiencies, and that, in the case of perfect detectors, it simplifies to a form which resembles the original Bell inequality \cite{Larsson}.  The CH-inequality, however, is a relation between the probabilities of having some events recorded in the experiment, and these probabilities can be easily modified to account for both: the inefficiencies of the detectors used and the noisy background \cite{Mermin2,Larsson,Eberhard}. Here we consider the Eberhard approach, where the CH-inequality is re-written as \cite{Eberhard,Garuccio}
\begin{eqnarray} \label{Eberhard}
P(\phi_2,\phi_3) &\leq &P(\tilde{\phi_1},\phi_3) + P(\phi_1,\phi_4) + P(\phi_2,\tilde{\phi_4}) \nonumber \\
& &+\frac{1-\eta}{\eta}[P(\phi_2) + P(\phi_3)],
\end{eqnarray} where $\eta$ is the detection efficiency of the measuring apparatuses. Let us now define $Q\equiv P(\phi_2,\phi_3) - P(\tilde{\phi_1},\phi_3) - P(\phi_1,\phi_4) - P(\phi_2,\tilde{\phi_4})$. When the value of $Q$ is positive, it represents the quantum violation of the CH-inequality. For such cases, the detection efficiency $\eta$ must be greater than a certain critical value to allow the violation of inequality (\ref{Eberhard}) without resorting to any supplementary assumption. This value is given by \cite{Garuccio}
\begin{equation} \label{etaGaru}
\eta_{crit} = \frac{P(\phi_2) + P(\phi_3)}{Q + P(\phi_2) + P(\phi_3)},
\end{equation} and one usually refers to $\eta_{crit}$ as the required efficiency for a detection loophole free Bell test, or also: threshold detection efficiency.

This expression emphasizes a behavior between the threshold detection efficiency and the quantum violation of the CH-inequality ($Q$), which is intuitive: it shows that the required efficiency is inversely proportional to the quantum CH-violation. Nevertheless, to completely understand the real behavior of $\eta_{crit}$, one has also to take into account its dependence on the probabilities $P(\phi_2)$ and $P(\phi_3)$ that appear on the right hand side of Eq~(\ref{etaGaru}). These probabilities are given by $P(\gamma) = tr(\rho_{red} P^{(k)}_{\gamma}) = \alpha^2 \sin^2(\gamma)+ \beta^2 \cos^2(\gamma)$, where $\rho_{red}$ is the reduced density operator of the composite system. It is clear, therefore, that the point of the minimal efficiency does not necessarily happen at the point where the quantum violation of the CH-inequality is maximal. As demonstrated numerically by Eberhard \cite{Eberhard} and later analytically by Larsson and Semitecolos \cite{Larsson2}, this minimum occurs for an almost product state and it has the value of $\eta_{min}=\frac{2}{3}$. The important property that Eq~(\ref{etaGaru}) emphasizes is that the value of $\eta_{crit}$ depends on the measurement settings considered, as is also the case with the CH-violation.

\begin{figure}[tbh]
\begin{center}
\rotatebox{270}{\includegraphics[width=0.35\textwidth]{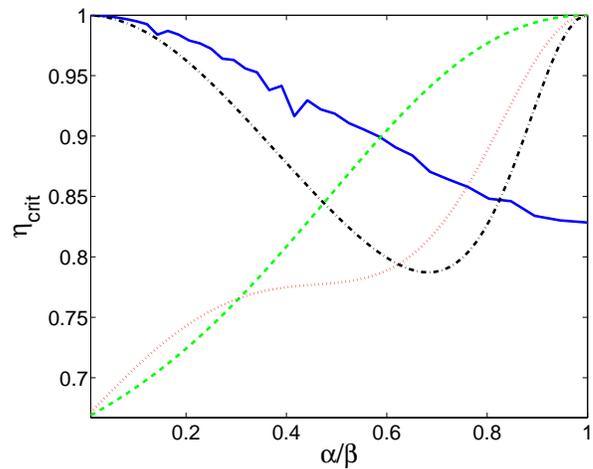}}
\end{center}
\vspace{-0.5cm}
\caption{ (Color online) The threshold detection efficiencies for the same measurement bases considered in Fig~\ref{CHVio}. The dashed curve (green curve) is obtained when one considers the Hardy measurements bases given in Eq~(\ref{Harbasis}). The dotted curve (red curve) and the dashed-dotted one (black curve) are obtained when the measurements are done with the experimental apparatuses in the orientations defined by the generalized Hardy measurement bases given by Eq.~(\ref{basis1}) and Eq.~(\ref{newbases}) with $n=1; m=7$ and $n=3; m=10$, respectively. The solid curve (blue curve) is obtained numerically with the measurement bases in the general form of Eq~(\ref{basis1}) and with a program maximizing the CH-violation.}  \label{eta1}
\end{figure}

In Fig.~\ref{eta1}, we now show the dependence of $\eta_{crit}$ with the same measurement bases considered in Fig.~\ref{CHVio}, to study the CH-inequality violation. The solid curve (blue curve) of Fig.~\ref{eta1} represents the required efficiency for a set of measurement bases that maximizes the CH-violation for each value of $\alpha/ \beta$. By analyzing Fig~\ref{CHVio} and Fig~\ref{eta1}, together, some interesting conclusions can be reached. (i) From the solid curves (blue curves) one can see that some measurement bases which generate the maximal CH-violation for a given value of $\alpha/ \beta$, do not necessarily allow the experiment to work with the lowest required efficiency for closing the detection loophole. In fact, they may even require the efficiencies of the detectors to be $100\%$ when the entanglement degree of the state considered in the test is low. (ii) From the dashed green curves one can see that to perform a Bell-test in the regime of lowest required efficiency, $\eta_{min}=\frac{2}{3}$, one can do the measurements in the directions defined by Hardy measurement bases [Eq.~(\ref{Harbasis})] and an entangled state for which $\alpha/\beta \approx 0.01$. Unfortunately, the corresponding violation of CH-inequality is very low at this point (See Fig~\ref{CHVio}), and the experimental errors involved will probably invalidate the experiment as a conclusive test of nonlocality, even if the detection efficiencies of the detectors are higher than $\eta_{min}$. (iii) The most interesting feature, however, is the behavior of the dashed-dotted curves (black curves) at Fig~\ref{CHVio} and Fig~\ref{eta1}. One can see from these curves that the minimum of the required efficiency $\eta_{crit}$, which is around $\alpha/ \beta = 0.7$, corresponds to a point where the CH-violation is very high. It is, therefore, legitimate to ask if there are measurement bases that can, at the same time and for each entangled state (for each value of $\alpha/\beta$), give the highest possible CH-violation, while requiring the lowest possible detection efficiency for a loophole free experiment. Hereafter we refer to these measurement bases as optimal. Moreover, it is also reasonable to assume that these optimal measurement bases could be generalizations of the bases used to drawn these black dashed-dotted curves.

\section{Optimal measurement bases}

To investigate this we considered more general measurement bases, which are defined through Eq~(\ref{basis1}) and by doing the sines and cosines functions as
\begin{eqnarray} \label{newbases2}
\sin{\phi_1} = \frac{\beta^{\frac{k_1}{2}}}{\sqrt{\alpha^{k_1} + \beta^{k_1}}}, \cos{\phi_1} = -\frac{\alpha^{\frac{k_1}{2}}}{\sqrt{\alpha^{k_1} + \beta^{k_1}}}, \nonumber \\
\sin{\phi_2} = \frac{\beta^{\frac{k_2}{2}}}{\sqrt{\alpha^{k_2} + \beta^{k_2}}}, \cos{\phi_2} = \frac{\alpha^{\frac{k_2}{2}}}{\sqrt{\alpha^{k_2} + \beta^{k_2}}}, \nonumber \\
\sin{\phi_3} = \frac{\beta^{\frac{k_3}{2}}}{\sqrt{\alpha^{k_3} + \beta^{k_3}}}, \cos{\phi_3} = -\frac{\alpha^{\frac{k_3}{2}}}{\sqrt{\alpha^{k_3} + \beta^{k_3}}}, \nonumber \\
\sin{\phi_4} = \frac{\beta^{\frac{k_4}{2}}}{\sqrt{\alpha^{k_4} + \beta^{k_4}}}, \cos{\phi_4} = \frac{\alpha^{\frac{k_4}{2}}}{\sqrt{\alpha^{k_4} + \beta^{k_4}}},
\end{eqnarray} and $\nu_{\phi_1} = \nu_{\phi_2} = \nu_{\phi_3} = \nu_{\phi_4} = 0$. This set of measurement bases can be seen as a generalization of the set defined in Eq~(\ref{newbases}), because here we do not force degeneracies on the measurements settings. That is, we do not require that the measurement orientations on the Alice side coincide or be symmetric to those used by Bob.

A new computer program was then written to minimize the threshold detection efficiency with these bases [Eq~(\ref{newbases2})], for each value of $\alpha/\beta$. The program performed an exhaustive search at the parameter space defined by the exponents $k_1$, $k_2$, $k_3$ and $k_4$, for each $\alpha/\beta$, with the values of these exponents varying from $1$ to $1024$. The program also recorded the CH-violation curve corresponding to the bases used. What turned out to be a surprise was the observation that the bases defined above in Eq~(\ref{newbases2}) are actually the bases which give, for most of the values of $\alpha/\beta$, the maximum possible CH-violation and require the lowest possible $\eta_{crit}$.

\begin{figure}[t]
\begin{center}
\rotatebox{270}{\includegraphics[width=0.35\textwidth]{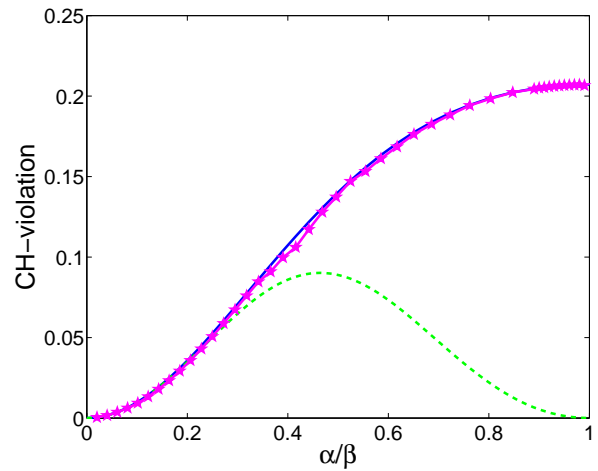}}
\end{center}
\vspace{-0.5cm}
\caption{(Color online) The CH-inequality violation of the generalized Hardy measurement bases defined by Eq~(\ref{newbases2}). The dashed curve (green curve) is obtained when one considers Hardy measurements bases given in Eq~(\ref{Harbasis}). The solid curve (blue curve) is the maximum allowed for the CH-violation. The star-marked curve (pink curve) is the CH-violation obtained when the measurements are done with the measuring apparatuses in the orientations defined by Hardy generalized measurement bases of Eq.~(\ref{newbases2}), and with the values of the coefficients $k_i$ given in Fig.~\ref{ks} and table~\ref{Tabla1}.} \label{CHVio2}
\end{figure}

\begin{figure}[tbh]
\begin{center}
\rotatebox{270}{\includegraphics[width=0.35\textwidth]{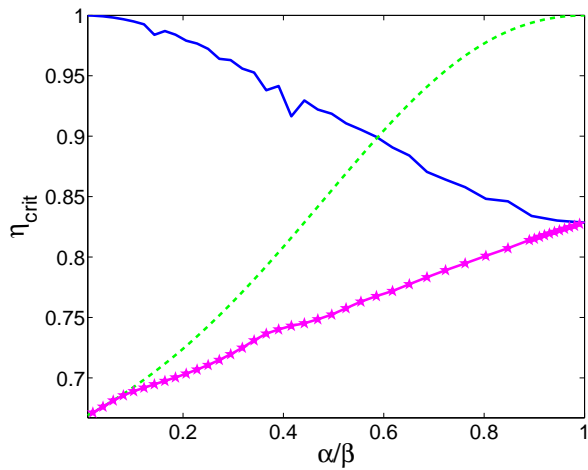}}
\end{center}
\vspace{-0.5cm}
\caption{(Color online) The threshold detection efficiency for the same measurement bases considered in Fig~\ref{CHVio2}. The dashed curve (green curve) is obtained when one considers Hardy measurements bases given in Eq~(\ref{Harbasis}). The solid curve (blue curve) is obtained numerically with the measurement bases in the general form of Eq~(\ref{basis1}) and with the program maximizing the CH-violation. The star-marked curve (pink curve) is obtained when the measurements are done with the measuring apparatuses in the orientations defined by Hardy generalized measurement bases of Eq.~(\ref{newbases2}), and with the values of the coefficients $k_i$ given in Fig.~\ref{ks} and table~\ref{Tabla1}.}  \label{eta2}
\end{figure}

\begin{figure}[tbh]
\begin{center}
\rotatebox{270}{\includegraphics[width=0.35\textwidth]{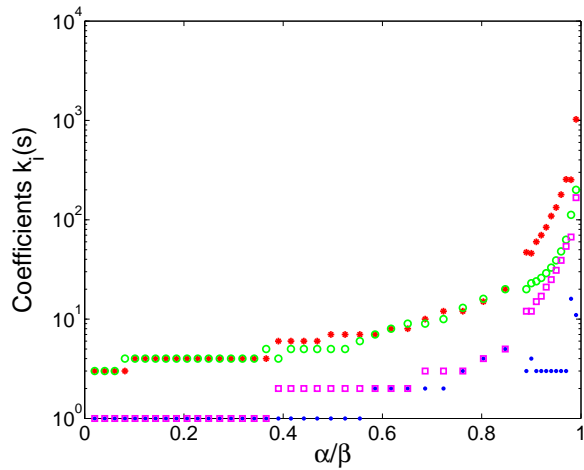}}
\end{center}
\vspace{-0.5cm}
\caption{(Color online) The coefficients $k_i$'s of the optimal measurement bases. These coefficients are plotted for each state considered in our calculations. The values considered of $k_1$ are shown with blue points. The values of $k_2$ are shown with red stars. The values of $k_3$ are shown as green circles and the values of $k_4$ are shown with pink squares. For higher values of $\alpha/\beta$ it is necessary to consider higher values of $k_i$(s), since at least one of the measurement projections tends to be always at the logical base $\{\ket{-^{(i)}},\ket{+^{(i)}}\}$.}  \label{ks}
\end{figure}

The curves obtained with these bases for the CH-violation and the required detection efficiency are given in Fig.~\ref{CHVio2} and Fig.~\ref{eta2}, respectively. The values of the coefficients $k_i$ are shown in Fig~\ref{ks} for each state considered in the calculations. Some of the values of these coefficients are given explicitly in table~\ref{Tabla1}. From Fig~\ref{ks} one can see that not necessarily the optimal measurement bases are degenerated or symmetric.

\begin{table*}[t]
\begin{ruledtabular}
\begin{tabular}{ccccccccc}
$\alpha/\beta$ &$k_1$ &$k_2$ &$k_3$ &$k_4$  &$\sin{\phi_1}$ &$\sin{\phi_2}$ &$\sin{\phi_3}$ &$\sin{\phi_4}$\\
\hline
$0.20$ &$1$ &$4$ &$4$ &$1$ &$0.91$ &$0.99$ &$0.99$  &$0.91$\\
$0.39$ &$1$ &$6$ &$4$ &$2$ &$0.84$ &$0.99$ &$0.98$ &$0.93$\\
$0.61$ &$2$ &$8$ &$8$ &$2$ &$0.85$ &$0.99$ &$0.99$ &$0.85$\\
$0.80$ &$4$ &$15$ &$16$ &$4$ &$0.84$ &$0.98$ &$0.98$ &$0.84$\\
$0.90$ &$4$ &$46$ &$23$ &$12$ &$0.77$ &$0.99$ &$0.95$ &$0.88$ \\
$0.95$ &$3$ &$133$ &$39$ &$31$ &$0.73$ &$0.99$ &$0.93$ &$0.91$ \\
$0.99$ &$11$ &$1024$ &$200$ &$167$ &$0.72$ &$0.99$ &$0.93$ &$0.91$
\end{tabular}
\end{ruledtabular}
\caption{Some of the calculated coefficients $k_i$'s of the optimal measurement bases [Eq~(\ref{newbases2})]. It is interesting to note that for every value of the parameter $\alpha/\beta$ the exponent $k_2$ is such that $\phi_2$ goes to $\pi/2$. This means that the measurement orientation $\ket{v^{(2)}}_{\phi}$ asymptotically reaches the logical state $\ket{+^{(2)}} $. The point $\alpha/\beta=0.99$ provides the optimal measurement bases for an almost maximally
entangled state.} \label{Tabla1}
\end{table*}

On Fig~\ref{CHVio2}, one can see that the measurement bases of  Eq. ~(\ref{newbases2}) can generate the maximal CH-violation curve. There are small discrepancies between the solid blue curve and the pink star-marked curve, but they can be made even smaller by increasing the time of computation on the generation of the pink star-marked curve. That is, by effectively increasing the parameter space. This can be done by considering the values of the coefficients $k_i$ in a larger interval range, or by considering the values of $\nu_{\phi_i} \neq 0$. Nevertheless, it is clear that for the majority of the entangled states considered, the bases given by Eq.~(\ref{newbases2}) were sufficient to generate the corresponding maximal CH-violation. For the other cases, these bases allow for almost maximal CH-violations.

In Fig.~\ref{eta2} there are three distinct curves. The solid blue curve and the dashed-green curves are the same curves discussed above. The star-marked curve (pink curve) is the curve for the required efficiency when considering our measurement bases of  Eq. ~(\ref{newbases2}) to define the orientation of the measuring apparatuses. This curve starts at the minimum efficiency of $\eta_{min} = \frac{2}{3}$ and slowly increases to $\eta_{crit} = 0.828$, when the composite system is a maximally entangled state. One can see that the required efficiency is much smaller in this case than it is when one chooses the bases that maximize the CH-violation without worrying with $\eta_{crit}$ (solid blue curve).

\begin{figure}[tbh]
\begin{center}
\rotatebox{270}{\includegraphics[width=0.35\textwidth]{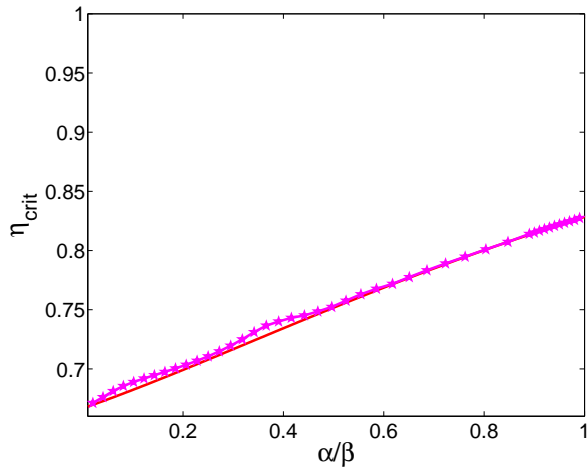}}
\end{center}
\vspace{-0.5cm}
\caption{(Color online) The lowest value allowed for $\eta_{crit}$ for a given value of $\alpha/\beta$. The solid curve (red curve) is obtained using the CG method. The star-marked curve (pink curve) is obtained when the measurements are done with the measuring apparatuses in the orientations defined by the optimal measurement bases given by Eq~(\ref{newbases2}) and the coefficients $k_i$'s showed in Fig.~\ref{ks} and table~\ref{Tabla1}.}  \label{eta3}
\end{figure}

Equation (\ref{newbases2}) is, in fact, a re-parametrization of the measurement bases of Eq.~(\ref{basis1}) as a function of $\alpha/\beta$. It is not clear that this parametrization can indeed generate the lowest curve possible for the required detection efficiency for a loophole free experiment. There could exist a curve for which the points between $66.7\%$ and $82.8\%$ would be below of those of the star-dotted pink curve of Fig~\ref{eta2}. To check this, we searched for the lowest curve possible for $\eta_{crit}$. We used again the the CG method to calculate the minimum of the required efficiency for each value of $\alpha/\beta$. This was done in the same way described before, with the program running with a large uniform sample of points in the parameter space defined by the variables $\{\phi_i\}$ and $\{\nu_i\}$. The required efficiencies obtained are shown with a solid red curve in Fig.~\ref{eta3}. At this figure there is also the curve of the required efficiencies of the generalized Hardy measurement bases given by Eq~(\ref{newbases2}) ($k_i$'s are given in Fig~\ref{ks} and table~\ref{Tabla1}). Again, there are small discrepancies between these curves. Whenever these curves superpose, one can say that the measurement bases given by Eq~(\ref{newbases2}) with the coefficients of Fig~\ref{ks}, are indeed the optimal measurement bases for Bell-type experiments based on the CH-inequality. At the other points we still have the curves close together, which shows that experimentally these bases are still a good choice for the tests of nonlocality based in the CH-inequality.

To demonstrate the relevance of the numerical results obtained, we
show in the Appendix the complexity of the analytical calculation of
the measurement bases that maximize the CH-violation for a certain
threshold detection efficiency. This calculation was presented by some
of us in \cite{Pino2011}, and here we review it. Starting at a given
value of $\eta_{crit}$, the calculation in the Appendix obtains the
optimal measurement bases and the corresponding value of
$\alpha/\beta$ analytically. The calculation, while possible, does not
add understanding (or accuracy) to the numerical procedure presented
here. In particular, the numerical method presented here easily adapts
to the case where the efficiencies are not symmetric, as opposed to
the analytical calculation.

\section{Conclusion}

In this work we investigated the relation between the violation of the CH-inequality, the efficiencies of the measuring apparatuses required for closing the detection loophole, and the measuring settings. It is well known that required efficiency decreases when one considers partially entangled states and that the use of these states, unfortunately, generates a reduction in the quantum violation of CH-inequality. Both quantities are dependent on the measurement settings used in the Bell test, and here we showed that there are measurement bases which allow for an optimal situation in this trade-off relation.

We have found the analytical form of these optimal measurement bases. The term optimal is used here to refer to the fact that these bases allow, for a given entangled state, the maximal possible CH-inequality violation while demanding the lowest possible detection efficiency for a loophole free experiment. These measurement bases are, therefore, of extreme relevance for future Bell experiments aimed to test nonlocality without assuming any supplementary assumption \cite{Science}. This results is specially important when one deals with systems of pair of entangled qubits, since the CH-inequality outperforms must of the known bipartite Bell inequalities \cite{BrunnerPLA}.

On this study we have considered the case for which the efficiencies of the measuring apparatuses involved in the test of nonlocality are equal. However, as discussed in Refs \cite{Cabello, Brunner}, the consideration of distinct efficiencies for the detectors may have an important practical consequence. They showed that when Alice's detectors are optimal, the Bob's ones may have an efficiency of $50\%$ for performing a conclusive Bell test based on the CH-inequality. The same analysis that we have done can be extended to the asymmetric Bell tests, and it is possible to demonstrate that the bases considered optimal in the symmetric case are also optimal for the asymmetrical tests.

\begin{acknowledgments}

The authors wish to thank to Ad\'{a}n Cabello for encouraging discussions. G.L. thanks to grants Milenio~P10-030-F and FONDECYT~11085055. C.S. acknowledges FONDECYT~1080383 and PFB08024. R.O.V. acknowledges the Brazilian agencies FAPEMIG and INCT-IQ.

\end{acknowledgments}

\appendix*
\section{Analytical derivation of the maximum violation of CH-inequality}
\label{An:analytical}

Here, we derive an expression for the maximal violation of the
CH-inequality given a certain threshold detection efficiency. This is
done by maximizing the eigenvalue of the CH-inequality operator.

For sake of simplicity, we use the projectors corresponding to the
Schmidt basis $\{\ket{\pm}^{(1)},\ket{\pm}^{(2)}\}$ \cite{Schimdt} as
the computational basis. We assume that the results of these
projectors are $a_1$ and $b_1$, so that
\begin{equation}
  \Pi_{a_1}=\left( \begin{array}{cc} 1 & 0 \\ 0 & 0 \end{array} \right)\otimes\mathbb I,
  \quad
  \Pi_{b_1}=\mathbb I\otimes\left( \begin{array}{cc} 1 & 0 \\ 0 & 0 \end{array} \right).
\label{eq:1}
\end{equation}
The rotations from $a_1$ to $a_0$ and from $b_1$ to $b_0$ can be
parameterized conveniently by
\begin{equation}
U_a=\left( \begin{array}{cc}  \sqrt{1-s} & \sqrt{s} \\  -\sqrt{s}  & \sqrt{1-s} \end{array} \right)\!,\
U_b=\left( \begin{array}{cc}  \sqrt{1-t}  & \sqrt{t}  \\  -\sqrt{t}  & \sqrt{1-t}  \end{array} \right).\
\label{eq:3}
\end{equation}
The $U_a$ rotation coincide with Eq. (\ref{basis1}) when
$\sqrt{s}=\cos{\phi}$ and $\nu_{\phi}=0$. We consider the
CH-inequality with the use of the following projectors
\begin{eqnarray}
    \Pi_{a_1b_1}&=&\Pi_{a_1}\otimes \Pi_{b_1}, \nonumber \\
    \Pi_{a_1b_0}&=&\Pi_{a_1}\otimes U_b^{-1}\Pi_{b_1}U_b, \nonumber \\
    \Pi_{a_0b_1}&=&U_a^{-1}\Pi_{a_1}U_a\otimes \Pi_{b_1}, \nonumber \\
    \Pi_{a_0b_0}&=&U_a^{-1}\Pi_{a_1}U_a\otimes U_b^{-1}\Pi_{b_1}U_b.
\label{eq:4}
\end{eqnarray}
The CH-inequality operator including efficiency, under the assumption
of independent errors at equal rate, is given by
\begin{equation}
    B =\eta^2(\Pi_{a_1b_1}+\Pi_{a_1b_0}+\Pi_{a_0b_1}-\Pi_{a_0b_0}) -\eta (\Pi_{a_1}+\Pi_{b_1}).
\end{equation}
It is worthwhile to remark that the quantum probabilities at the
expected value of the CH-inequality operator on state (\ref{Psi}) is
equivalent to Eq.~(\ref{etaGaru}), see \cite{Larsson2}. The
eigenvalues of this operator are the solutions of
\begin{eqnarray}
    && st\eta^5\left(-st\eta^3 +(s+t)\eta(2\eta-1)-3\eta+2\right) \nonumber \\
    &&\quad+2(\eta-1)\eta^3\left(st(\eta^2-\eta)-1\right)\lambda \nonumber \\
    &&\quad-\eta^2(4\eta-5)\lambda^2-2\eta(\eta-2)\lambda^3+\lambda^4=0.
    \label{Eq:CH_st}
\end{eqnarray}
Local Realism bounds the eigenvalues below zero, so any positive
eigenvalues will give a violation. Seeking a maximum violation, we
need to find the parameter values of $s$ and $t$ that gives this
maximum. We can also view this as finding $s+t$ and $st$ that gives
the maximum. $s+t$ only occurs in the constant term in the polynomial
so that, for a given value of $st$, the maximum $\lambda$ is obtained
when $s+t$ is minimal, i.e., when $s=t$. This reduces the unknowns,
and we have
\begin{eqnarray}
  && t^2\eta^5\left(-t^2\eta^3 + 2t\eta(2\eta-1)-3\eta+2\right) \nonumber \\
  &&\quad+2(\eta-1)\eta^3\left(t^2(\eta^2-\eta)-1\right)\lambda \nonumber \\
  &&\quad -\eta^2(4\eta-5)\lambda^2-2\eta(\eta-2)\lambda^3+\lambda^4=0.
  \label{Eq:CH_tt}
\end{eqnarray}
The singlet state $|+^{(1)},-^{(2)}\rangle-|-^{(1)},+^{(2)}\rangle$ is
an eigenvector of the operator $B$ with the eigenvalue
$\lambda_4=\eta^2t-\eta$, which is always negative. The remaining
three eigenvalues can be obtained by solving the third-degree equation
\begin{eqnarray}
    &&\frac{\lambda^3}{\eta^3}+\left(\eta(t-2)+3\right)\frac{\lambda^2}{\eta^2}\nonumber \\
   &&\quad +\left(\eta^2(t^2-2t)+2\eta(t-1)+2\right)\frac\lambda\eta \nonumber \\
    &&\quad + \eta^3t^3-3\eta^2 t^2+2\eta t^2=0.
\label{eq:2}
\end{eqnarray}
Solving the above equation using the trigonometric method gives us
\begin{widetext}
\begin{eqnarray}
    \lambda_1&=&-\frac{1}{3}\eta(3+(-2+t)\eta)+\frac{2}{3}
    \eta\sqrt{3-6\eta+(4+2t-2t^2)\eta^2}\nonumber\\
    &&\times \cos\left[\frac{1}{3}
      \arccos\left[\frac{ \eta\Big(9-18\eta+8\eta^2 -10t^3\eta^2+ 3t\big(3-6\eta+2\eta^2\big) -3t^2\big(9-18\eta+4\eta^2\big)\Big)}
        {\sqrt{(3-2\eta(3+(-2+t)(1+t)\eta))^3} } \right]\right].
\label{eq:5}
\end{eqnarray}
The other solutions can be obtained by adding $2\pi/3$ and $4\pi/3$ to
the $\arccos$ angle. These will be lower than $\lambda_1$ above. The
next step is to use the equation (\ref{eq:2}) as an implicit
definition of $\lambda_1$ as a function of $t$, and do implicit
differentiation with respect to $t$. Since we are seeking the maximum
value, $\lambda'(t)=0$, and the resulting second-degree equation can
be substituted back into (\ref{eq:2}), to obtain
\begin{equation}
\lambda = \frac{\eta}{2(\eta-1)^2}(2t^2\eta^3-3t\eta(2\eta-1)+3\eta-2), \label{eq:8}
\end{equation}
This can now be used in the second-degree equation to give a
fourth-degree polynomial equation for $t$ as
\begin{eqnarray}
    4\eta^6t^4 +4\eta^4\left(2\eta^2-10\eta+5\right)t^3
    +\eta^2\left(4\eta^4-48\eta^3+156\eta^2-132\eta+33\right)t^2 && \nonumber \\
    +2(2\eta-1)^2\left(5 \eta ^2-16\eta+8\right)t
    -(\eta-2)(2\eta-1)^2(3\eta-2)&=&0.
\label{eq:9}
\end{eqnarray}
\end{widetext} In principle, the solution of this equation gives us
the parameters of the rotation from $a_1$ to $a_0$ and from $b_1$ to
$b_0$ that determine the maximum violation of the CH-inequality
operator including efficiency. The eigenvectors will give us the
optimal quantum state. However, the analytic solution of the above
expression can be very long.

The asymmetric case ($\eta_A\neq\eta_B$) can be treated in the same
manner, but naturally, the procedure will be much more
complicated. The main complication is that the singlet state will not
be an eigenvector anymore, and that the polynomial equation for $t$
will be of higher degree than four. Here, one may have to resort to
numerical solution of the equation, in which case any benefit from
solving the system analytically disappears.

\end{document}